\documentclass[%
 aip,
 amsmath,amssymb,
 reprint,%
]{revtex4-1}

\usepackage{graphicx}
\usepackage{dcolumn}
\usepackage{bm}

\usepackage[utf8]{inputenc}
\usepackage[T1]{fontenc}
\usepackage{mathptmx}

\begin{document}

\preprint{AIP/123-QED}

\title[Magnetized Plasma Sheath in the Presence of Negative Ions]{Magnetized Plasma Sheath in the Presence of Negative Ions}

\affiliation{Centre of Plasma Physics, Institute for Plasma Research, Sonapur-782402, Kamrup (M), Assam, India}
\affiliation{Department of Physics, University of Oslo, PO Box 1048 Blindern, NO-0316 Oslo, Norway}
\affiliation{Lovely Professional University, Jalandhar Delhi-GT Road, Phagwara-144411 Punjab, India}

\author{R. Paul$^{1}$, S. Adhikari$^{2}$, R. Moulick$^{3}$, S. S. Kausik$^{1,}$}
\email{kausikss@rediffmail.com}
\author{B. K. Saikia$^{1}$}
\date{\today}

\begin{abstract}
 The sheath formation in a weakly magnetized collisionless electronegative plasma consisting of electrons, negative and positive ions has been numerically investigated using the hydrodynamic equations. The electrons and negative ions are assumed to follow Boltzmann relation. A sheath formation criterion has been analytically derived. The paper focuses on studying the sheath structure by varying the electronegativity. It has been observed that the presence of negative ions has a substantial effect on the sheath structure. The observations made in the present work have profound significance on processing plasmas, especially in the semiconductor industry as well as in fusion studies. 
\end{abstract}

\maketitle
\section{\label{intro}Introduction}
The physics of sheath has been intriguing to the researchers for several decades as it harbors diverse physical phenomena. The sheath structure, both in an electrostatic scenario as well as in the magnetized case, are widely studied\cite{moulick2019}. The knowledge of sheath helps in understanding the interaction of plasmas with different material surfaces\cite{khoramabadi}. Besides being one of the oldest topics in plasma, it finds its application in plasma technology, fusion research, and many more \cite{Riemann_1991}.  In fusion devices such as tokamaks, the magnetic field often makes an angle with the boundary wall. With the ongoing developments in fusion research, the study of the sheath in the oblique or tilted magnetic field has gained utmost importance\cite{doi:10.1063/1.863955,doi:10.1063/1.870800,Pc2012,moulick2019}.\\

It is well known that in the absence of the magnetic field, a non-neutral region appears near the wall. It consists of two subregions namely, the pre-sheath and the sheath region. The pre-sheath is located near the bulk plasma, whereas the sheath lies closer to the wall. The positive ion flux increases in the pre-sheath region and remains constant in the sheath. The sheath so formed is known as the Debye sheath. Usually the sheath begins when the ions satisfy the Bohm criterion given by $$ v_i \geq c_s = \sqrt{\frac{ T_e}{m_i }}$$\\ where $v_i$ and $ c_s$ are the positive ion velocity at the sheath edge and the Bohm speed, $T_e$ and $m_i$ are the electron temperature and positive ion mass, respectively. On the other hand, in the presence of the magnetic field, adjacent to the Debye sheath, another layer forms known as the magnetic presheath or Chodura sheath. If plasma contains negative ions as one of its components, then the sheath structure has been observed to be modified\cite{doi:10.1063/1.3077304,hatami}, and such plasmas are termed as electronegative plasmas. For such plasma, the sheath potential was studied by Thompson and Boyd in 1950s\cite{doi:10.1098/rspa.1959.0140}. In an electronegative  plasma, the Bohm speed  has been modified as$$ v_i \geq c_s =\sqrt{\frac{ (n_-+n_e ) T_-}{n_- T_e+n_e T_- }}$$  where  $T_-$ is the temperature of the negative ions and $n_e$, $n_-$ are the density of the electrons and negative ions respectively\cite{hatami,allen}.   \\ 

Electronegative gases and plasmas have been seen to gather a lot of attention in recent times as it has applications in different fields, such as surface processing, atmospheric science, environmental studies,  and many others \cite{Stoffels,palop}. 
These electronegative plasmas are also used as a negative ion beam generators, which are used as neutral beams in fusion reactors\cite{Stoffels,Dubois}. The presence of the negative ions is also said to improve the performance of dry etching\cite{Ohtake}, which enables in producing integrated circuits in bulk\cite{doi:10.1063/1.3310837}. In low-pressure plasmas, the negative ions are usually created by volume production\cite{bacal,BECHU}, but the negative ions are also produced by surface production as well \cite{Dubois}. Recently, Kakati \textit{et al}.\cite{kakati} have developed a novel process of producing negative hydrogen ions ($H^-$) by cesium coated tungsten dust. They have reported of producing the $H^-$ ions by the surface process as it has an advantage over the two-step volume process \cite{kakati}.\\

One of the important parameters for defining negative ion plasmas is the electronegativity ($\alpha=\frac{n_-}{n_e}$). Many researchers have studied electronegative plasmas in the presence of an external magnetic field. Yasserian and Aslaninejad \cite{doi:10.1063/1.3310837} studied the presheath-sheath in magnetized electronegative discharge in the presence of magnetic field and elastic collisions including ion-neutral collisions. Hatami \textit{et al.}\cite{hatami} studied the sheath in presence of negative ions in a multicomponent plasma considering two species of positive ions and electrons neglecting the effect of collisions, ionization and recombination.  Franklin and Snell \cite{franklin}, on the other hand, studied the sheath structure in the presence of an external magnetic field but with only one species of positive and negative ions each. Besides the fluid approach, kinetic approach has also been considered in studying the electronegative plasmas \cite{oudini}.\\

 In addition to the above mentioned studies, the effect of negative ion temperature and electronegativity seems to have a notable influence on the sheath structure which has not been explored. In the present work, a collionless, low-temperature electronegative plasma has been investigated using a single fluid approach. The current work focusses on studying the effect of varied electronegativity on the sheath structure . It also emphasizes the impact of the negative ion temperature in understanding the dynamics involved in the formation of the sheath. \\

The paper has been divided into the following sections, Sec.\ref{MODEL AND BASIC EQUATIONS} discuss the model and the set of equations used in modeling the system. Sec. \ref{NUMERICAL EXECUTION} and Sec. \ref{RESULTS AND DISCUSSIONS 
} analyze the numerical findings and a brief conclusion has been presented in Sec.\ref{CONCLUSIONS}.

\section{\label{MODEL AND BASIC EQUATIONS}MODEL AND BASIC EQUATIONS}
\begin{figure}
    \centering
    \includegraphics[width=8.5cm]{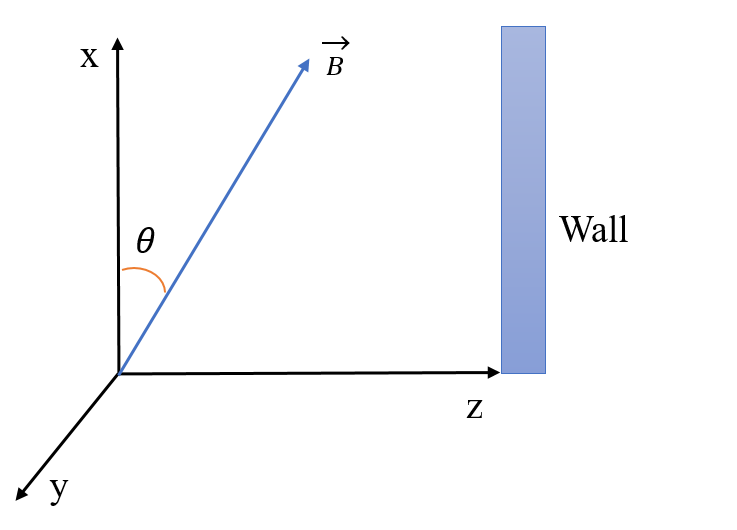}
    \caption{Geometry of the system.}
    \label{fig:geometry}
\end{figure}

A multi-component magnetized plasma has been considered comprising of electrons, positive ions, and negative ions. The plasma is considered to be in contact with a planar wall, as shown in Fig.\ref{fig:geometry}. An oblique magnetic field \textbf{B} is considered which lies in the x-z plane with an angle of inclination, $\theta$  to the x-axis perpendicular to the wall. The surface is parallel to the x-axis. The plasma is assumed to be in steady-state, and the ion-neutral collisions have been neglected.\\
The electrons and negative ions can be described by the Boltzmann relation, which is given by \\
\begin{equation}
    n_e=n_{eo}\exp{\frac{e\phi}{kT_e}} ,
  \end{equation}  
   \begin{equation}
      n_-=n_{-o}\exp{\frac{e\phi}{kT_-}}  .  
   \end{equation}
	                                                                
where $T_e$ and $T_-$ represents the temperature of the electrons and the negative ions, respectively , $\phi$ is the electric potential,  $n_{eo}$ and $n_{-o}$ represents the density of the electrons and negative ions, respectively. An isotropic medium has been considered, where all species have uniform temperature in all directions.\\

The electrons are chosen to be Boltzmannian due to their negligible mass.Apart from this approach, the electrons being described by Boltzmann relation can be justified when the pressure gradient balances the electric force\cite{beilis}. This relation is a consequence of the absence of the magnetic field, collision, and particle generation in the momentum balance equation for electrons. In absence of the other factors, the presence of the magnetic field alone may be sufficient to cause the deviation from the Boltzmann behavior \cite{R}. However, in case the electrons are strongly magnetized, the fluid velocity is of the electrons is small as compared to the thermal velocity, and this retains the Boltzmann behavior.  According to Chen\cite{Chen}, the Maxwellian distribution can be considered in the presence of a dc magnetic field as the Lorentz force is perpendicular to the velocity and thus unable to impart energy to any electron.\\

Negative ions as well have been described by the Boltzmann relation. The electrons and the negative ions are pushed by the electric field back towards the center of the discharge in the presheath region, which decreases the velocities of the electrons and the negative ions. The magnetic field force has a negligible effect on the negative ions as compared to the impact on the positive ions\cite{doi:10.1063/1.3077304}. This aids in the balance between the drift and the diffusion, thereby validates the description of negative ions by Boltzmann relation.\\

The cold ions are governed by the hydrodynamic equations given by
\begin{equation}
    m_i n_i (\textbf{v} .{\nabla})\textbf{v}   =  - n_i e (\textbf{E} + \textbf{v} \times \textbf{B} ) - m_i S_i\textbf{v} 
\end{equation}
The continuity equation is given by  \\
\begin{equation}
    {\nabla}.(n_i\textbf{v})= S_i                                                                         
\end{equation}
where
                   $$ S_i    = n_e \mathbb{Z}   $$

Here \textbf{B} is the magnetic field intensity, $ \mathbb{Z}$   is the ionization rate which is assumed to be constant, $m_i$,$v$ are the mass and velocity of the positive ions respectively.\\

Considering the 1D-3V representation of the system, the equations can be resolved into components along the x, y and z directions as\\
\begin{equation}
    v_z  \frac{dv_x}{dz} = \omega_z v_y - \frac{n_e\mathbb{Z}}{n_i}v_x  ,                     
\end{equation}
\begin{equation}
    v_z  \frac{dv_y}{dz} = (\omega_x v_z -\omega_z v_x) - \frac{n_e\mathbb{Z}}{n_i}v_y ,                       
\end{equation}    
\begin{equation}
    v_z  \frac{dv_z}{dz} =-\frac{e}{m_i}\frac{d\phi}{dz} -\omega_x v_y  - \frac{n_e\mathbb{Z}}{n_i}v_z ,   
\end{equation}    
\begin{equation}
\frac{d(n_iv_z)}{dz} =n_e\mathbb{Z} .             
\end{equation}
The electrostatic potential is governed by the Poisson equation
\begin{equation}
             \frac{d^2\phi}{dz^2} =  -\frac{ e}{\epsilon_o }(n_i -n_e - n_- ) .     
\end{equation}
To solve the above set of equations, it is necessary to normalize the quantities by the following dimensionless parameters
$$ u= \frac{v_x}{c_s}  ,~~ v =\frac{v_y}{c_s}  ,~~ w = \frac{v_z}{c_s},$$                                                       
     $$        \gamma_x =\frac{\omega_x\lambda_{ni}}{c_s } ,~~ \gamma_z =\frac{\omega_z\lambda_{ni}}{c_s } ,~~  
     \beta = \frac{\mathbb{Z}\lambda_{ni}}{c_s },$$
           $$     \xi = \frac{z}{\lambda_{ni}}, ~~ \eta =-\frac{e\phi}{kT_e} ,~~ \lambda_{ni}=\frac{c_s}{\mathbb{Z}} ,$$   
    $$    N_i = \frac{n_i}{n_{io} } ,~~ N_e = \frac{n_e}{n_{io} }  ,~~ N_- = \frac{n_-}{n_{io}}, ~~\alpha=\frac{n_{-o}}{n_{eo}} , $$ 
	$$     \delta=\frac{n_{eo}}{n_{io} }  ,~~ 1 - \delta = \frac{n_{-o}}{n_{io} } ,\tau_-=\frac{T_e}{T_n}.$$
	
	where ,
$$	c_s    = \sqrt{\frac{kT_e}{m_i}},  ~~ \lambda _D=\sqrt{\frac{\epsilon kT_e}{n_{eo}e^2}} $$ 
$$	\omega_x  =  \frac{eB_x}{m_i} =\frac{eB_o\cos{\theta}}{m_i}  $$    
           $$  \omega_z   = \frac{eB_z}{m_i} =  \frac{eB_o\sin{\theta}}{m_i}$$
         $\omega_x $ and $ \omega_z $ are the respective ion- cyclotron frequencies along the x and z directions.\\
Here the equations are scaled in the ionization scale as the pre sheath region is also considered.\\
The normalized equations are given by 
\begin{equation}
   \frac{du}{d\xi}  =    \gamma_z \left(\frac{v}{w}\right) - \beta \frac{N_e}{N_i } \frac{u}{w}    ,                
\end{equation}
\begin{equation}
   \frac{dv}{d\xi}  =   \gamma_x- \gamma_z \left(\frac{u}{w}\right) - \beta \frac{N_e}{N_i } \frac{v}{w} ,                   
\end{equation}
\begin{equation}
   \frac{dw}{d\xi}  =  \frac{1}{w}\frac{d\eta}{d\xi} -\gamma_x \left(\frac{v}{w}\right) - \beta\frac{N_e}{N_i } ,
   \end{equation}
  \begin{equation}
   \frac{dN_i}{d\xi}  =  2\beta\frac{N_e}{w}-\frac{N_i}{w^2}\frac{d\eta}{d\xi} -\gamma_x\left( \frac{ N_i}{w^2}\right)v, 
   \end{equation}
      \begin{equation}
   \frac{d^2\eta}{d\xi^2}   =  a_o(N_i-\delta N_e - (1 -\delta)N_-)  ,
   \end{equation}
   where  $$ a _o  =\left(\frac{\lambda_{ni}}{\lambda_D}\right)^2 . $$
   The normalized electron and negative ion densities are given by
   \begin{equation}
       N_e = \exp(-\eta),~~ N_- = \exp(-\eta \tau_-).
   \end{equation}
  \subsection{The sheath formation criterion}
 Using Poisson’s equation, the space charge variation can be written as\\
   $$\frac{d\sigma}{dz}=\frac{d(n_i-n_e-n_-)}{dz} $$
   \begin{equation}
     or,  \frac{d\sigma}{dz}= \frac{n_i}{v_z} \left[\Omega-\frac{eE}{m_iv_z}+\frac{v_y}{v_z}\omega_x\right]
        + \frac{n_eeE}{T_e}  + \frac{n_-eE}{T_e}\tau_-
   \end{equation}
     where     $$   \Omega = \frac{2n_e\mathbb{Z}}{n_i} $$ 
        For sheath formation,\\
        $$\frac{d\sigma}{dz}> 0$$
        $$or, \frac{n_i}{v_z} \left[\Omega-\frac{eE}{m_iv_z}+\frac{v_y}{v_z}\omega_x\right]
        + \frac{n_eeE}{T_e}  + \frac{n_-eE}{T_e}\tau_- > 0$$
        $$or, \frac{eE}{m_iv_z} \frac{n_i}{v_z }\left[\frac{n_e v_z^2 m_i}{n_i T_e }+\frac{n_- v_z^2m_i\tau_-}{n_i T_e }-1\right]+  \frac{n_i}{v_z }\left[\Omega+\frac{v_y}{v_z }\omega_x \right]     > 0$$
   $$or,    
   \frac{eE}{m_iv_z} \frac{n_i}{v_z }\left[\frac{ n_e v_z^2}{n_i (T_e/m_i) }+\frac{n_- v_z^2\tau_-}{n_i (T_e/m_i) }-1\right]+  \frac{n_i}{v_z }\left[\Omega+\frac{v_y}{v_z }\omega_x \right] > 0$$
   \begin{equation}
     or,\frac{eE}{m_iv_z} \frac{n_i}{v_z }\left[\frac{ v_z^2}{(n_i/n_e) v_B^2}+\frac{ v_z^2\tau_-}{(n_i/n_-)v_B^2}-1\right]+  \frac{n_i}{v_z }\left[\Omega+\frac{v_y}{v_z }\omega_x \right] > 0   
   \end{equation}
As $\frac{n_i}{v_z}\ne 0$,  therefore , equation (19) can be rewritten as 
$$ \frac{eE}{m_iv_z}\left[\frac{ v_z^2}{(n_i/n_e )v_B^2}+\frac{ v_z^2\tau_-}{(n_i/n_-) v_B^2}-1\right]+  \frac{n_i}{v_z }\left[\Omega+\frac{v_y}{v_z }\omega_x \right] > 0  $$

$$ or, ~~ \frac{eE}{m_iv_z} \left[ 1- \frac{ v_z^2}{v_c^2} - \frac{ v_z^2 \tau_-}{v_n^2} \right]    
 <  \Omega \left[ 1+\frac{v_y}{v_z } \frac{\omega_x}{\Omega} \right] $$
 \begin{equation}
     or,~~\left[1-v_z^2\left(\frac{1}{v_c^2}+\frac{\tau_-}{v_n^2}\right)\right]  < \frac{v_z}{v_\Omega}\left[1 + \frac{v_y}{v_z} \frac{\omega_x}{\Omega}\right] 
 \end{equation}
 where $$  v_\Omega= \frac{eE}{m_i\Omega}  ,  v_c=v_B \sqrt{\frac{n_i}{n_e}} ,v_n= v_B\sqrt{\frac{n_i}{n_-}}. $$
Equation (18) represents the general condition for the sheath formation. It is seen that the sheath criterion depends on the ion-cyclotron frequency, the respective densities of positive ions, electrons, and negative ions and also on the temperature ratio between the electrons and the negative ions.\\
 The wall is assumed to be floating, thus the flux balance has been given by
\begin{equation}
         n_i v_z = \frac{1}{4} n_e\bar{c}.                 
\end{equation}
 where $\bar{c} = \sqrt{\frac{8T_e}{\pi m_e}}$  , which is the random velocity of the electron at the wall. This flux balance condition is acquired by the equating the ion drift and the random thermal motion of the electrons\cite{moulick2019,valentini,forrest}.

\section{\label{NUMERICAL EXECUTION}NUMERICAL EXECUTION}
Equations (10) - (14) are a set of ordinary differential equations which can be solved using the Runge-Kutta fourth order method.These are initial valued problem and the values of the dependent variable must be supplied at the mid-plane (initial point) of the plasma. On considering $\xi$ = 0 as the initial point,the quantities such as velocities, densities becomes zero. Hence, a point close to $\xi$ = 0 is considered where the initial values are obtained using Taylor series expansion \cite{valentini,crespo,moulick2019}.\\
The series may be constructed as
$$V_i = V_{i1}\xi + V_{i2}\xi^3 + ……………$$
$$\eta=  \eta_1  \xi^2 + \eta_2  \xi^4 + ……………$$
where N, V, $\eta$ represents the density, velocity, and potential, respectively for ions. \\
The first order coefficient of the above 
$$\eta_1  = 0 ,$$
$$u_1  =  \frac{\gamma_x \gamma_z N_{eo}}{4N_{eo}^2+\gamma_z^2 },$$ 
$$ v_1   =  \frac{2\gamma_x N_{eo}^2}{4N_{eo}^2+\gamma_z^2}, $$ 
$$ w_1  = \frac{N_{eo}}{N_{io}},  $$ 
$$  N_{i0}  = 1. $$
These coefficients become the initial values and are used to obtain the solution of the above differential equations. The initial point is considered nearby $\xi$= 0, as small as possible, of the order of $10^{-4}$or $10^{-5}$.\\
The default parameters chosen for simulation are\\
$T_e = 1eV; m_i= 1 AMU ; \mathbb{Z} = 1 \times 10^ 5 s^{-1}; \\$
$n_o = 1 \times 10^{18} m^{-3};B_0 = 0.35T;$
$\theta=20^{0};\tau_- =25.$ \\
Here  $H_2$ plasma is considered, the positive, as well as the negative ions, are that of $H_2$. \\
The differential equations are thus solved using MATLAB routine ODE45.

\section{\label{RESULTS AND DISCUSSIONS }RESULTS AND DISCUSSIONS}

\begin{figure}
    \centering
    \includegraphics[width = 8.5 cm]{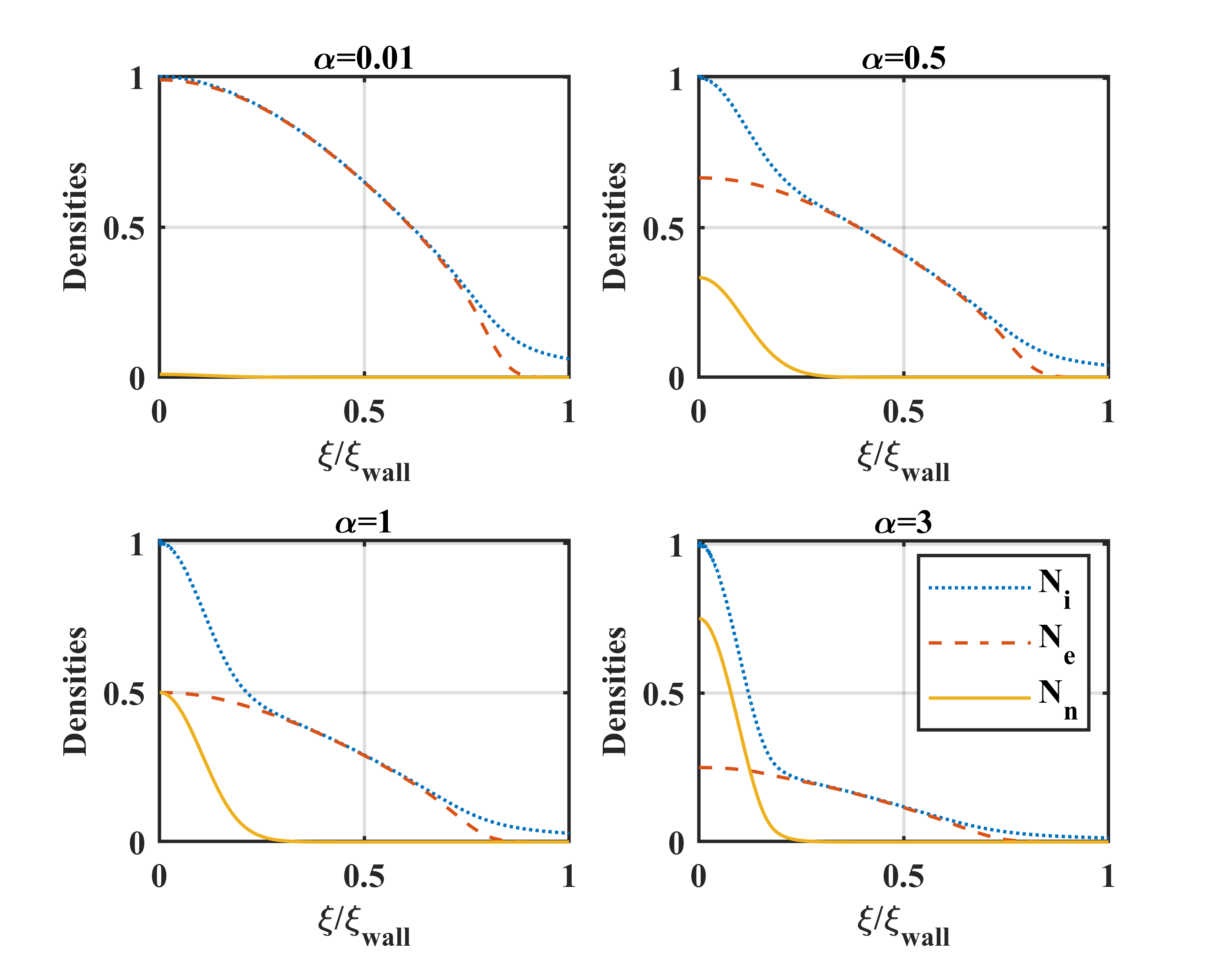}
    \caption{Normalized density profile of all the species along the normalized $z$ axis.}
    \label{fig:density_1}
\end{figure}

 \begin{figure}
    \centering
    \includegraphics[width = 8.5 cm]{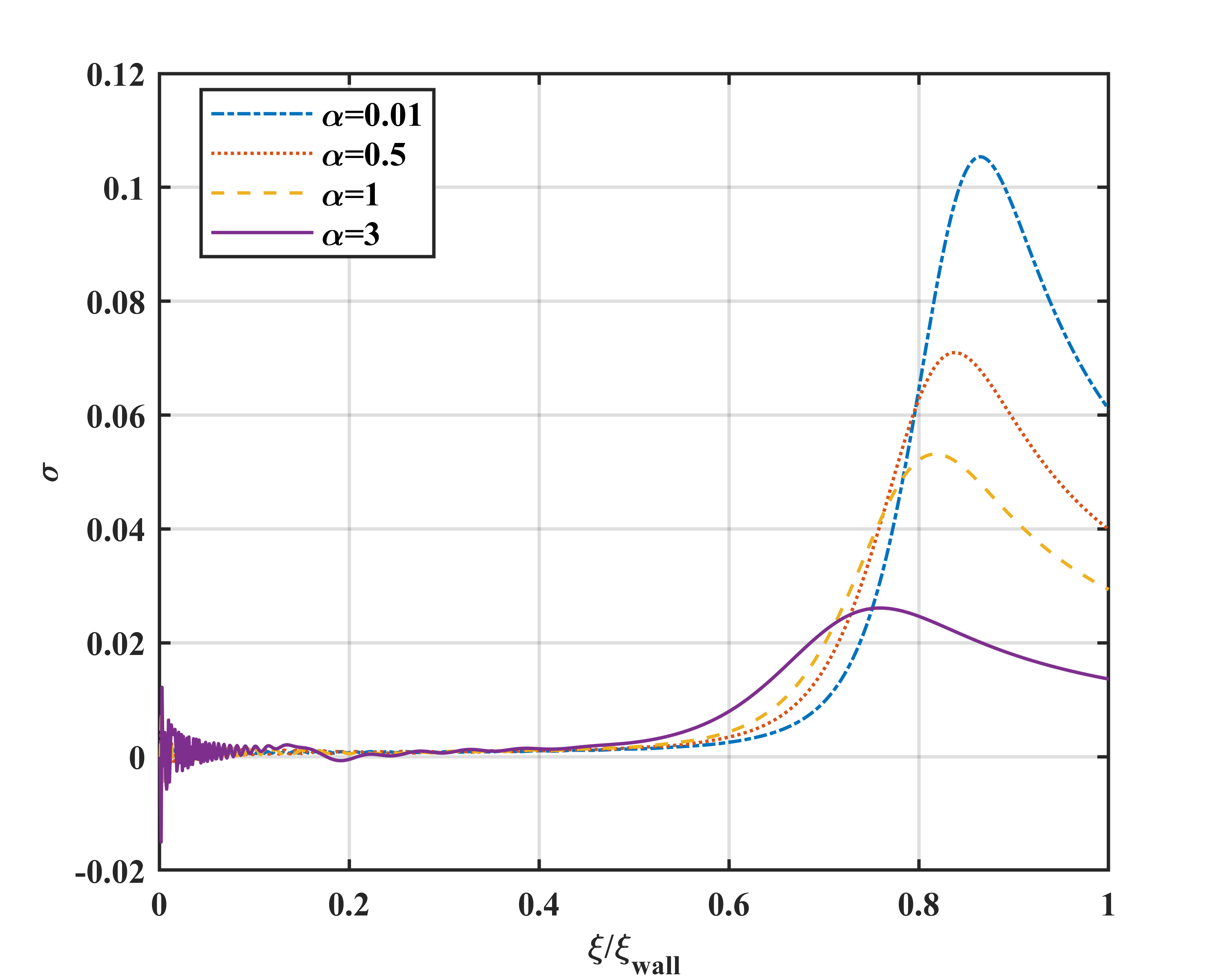}
    \caption{Space charge variation along normalized $z$ axis.}
    \label{fig:space charge} 
\end{figure}

\begin{figure}
    \centering
    \includegraphics[width = 8.5 cm]{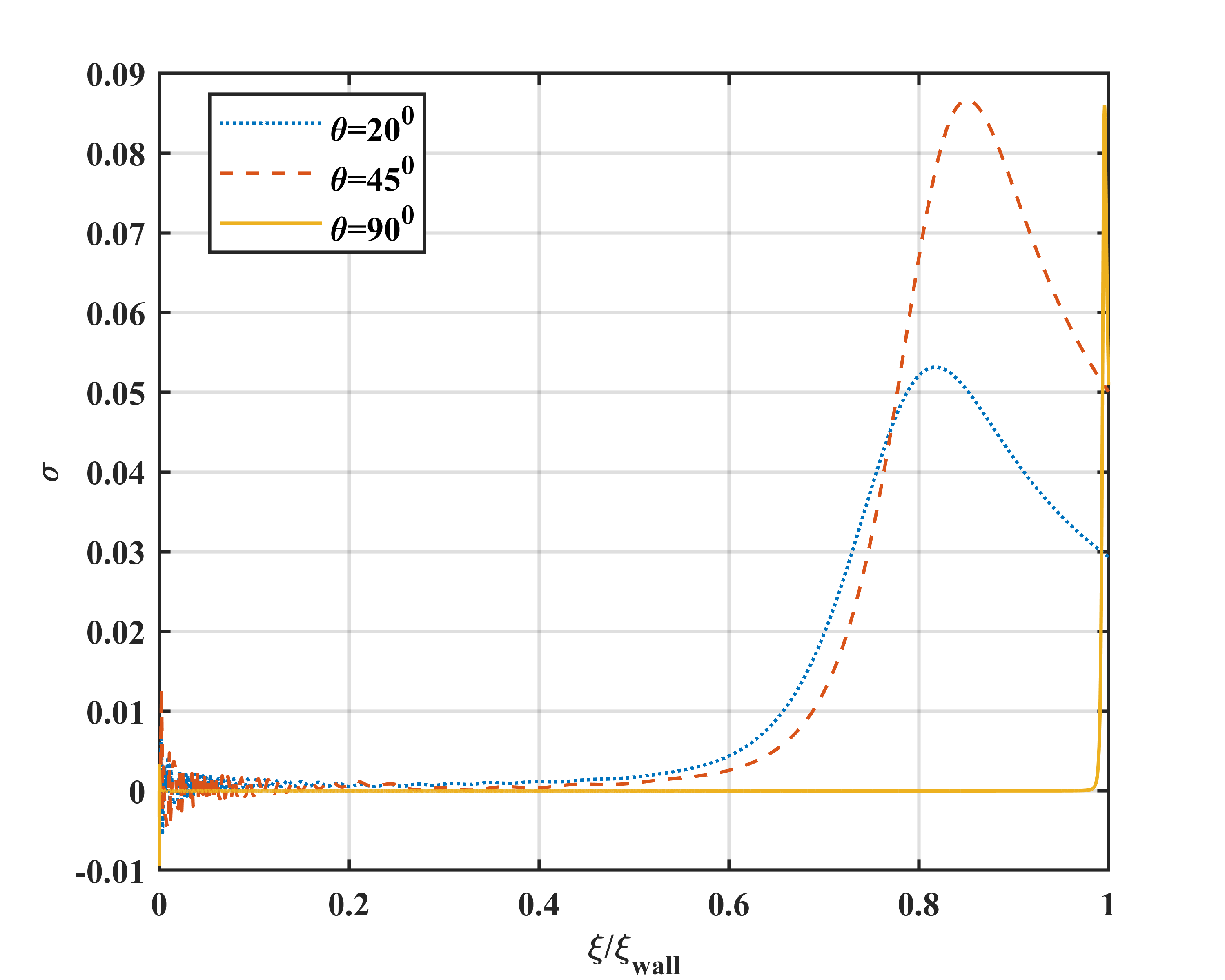}
    \caption{Space charge variation along normalized $z$ axis for different $\theta$ for $\alpha=1$}
    \label{fig:space_charge_2} 
\end{figure}

\begin{figure}
    \centering
    \includegraphics[width = 8.5 cm]{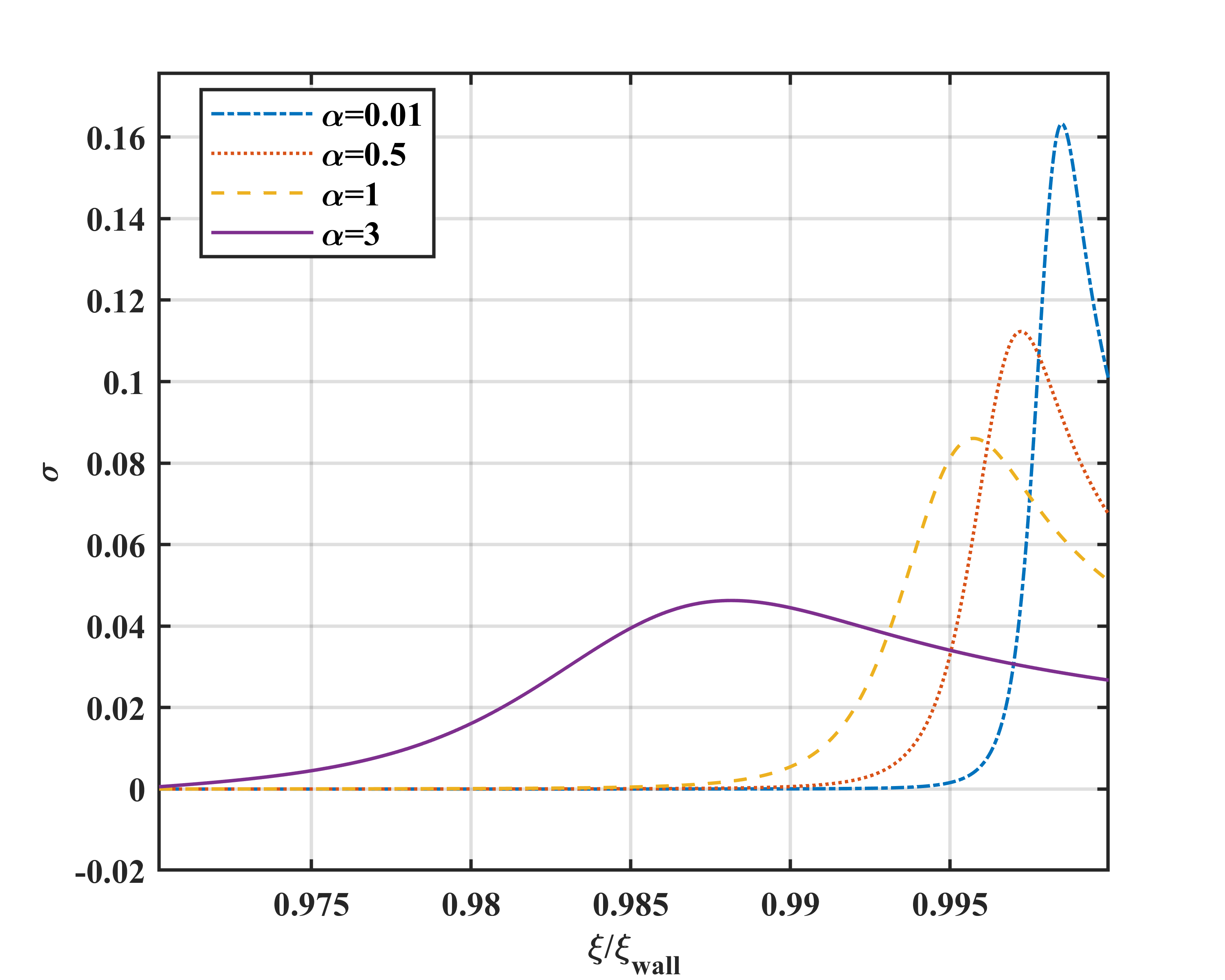}
\caption{Space charge variation along normalized $z$ axis for different $\alpha$ for $\theta=90^{\circ}$}
    \label{fig:space_charge_90} 
\end{figure}

\begin{figure}
    \centering
    \includegraphics[width = 8.5 cm]{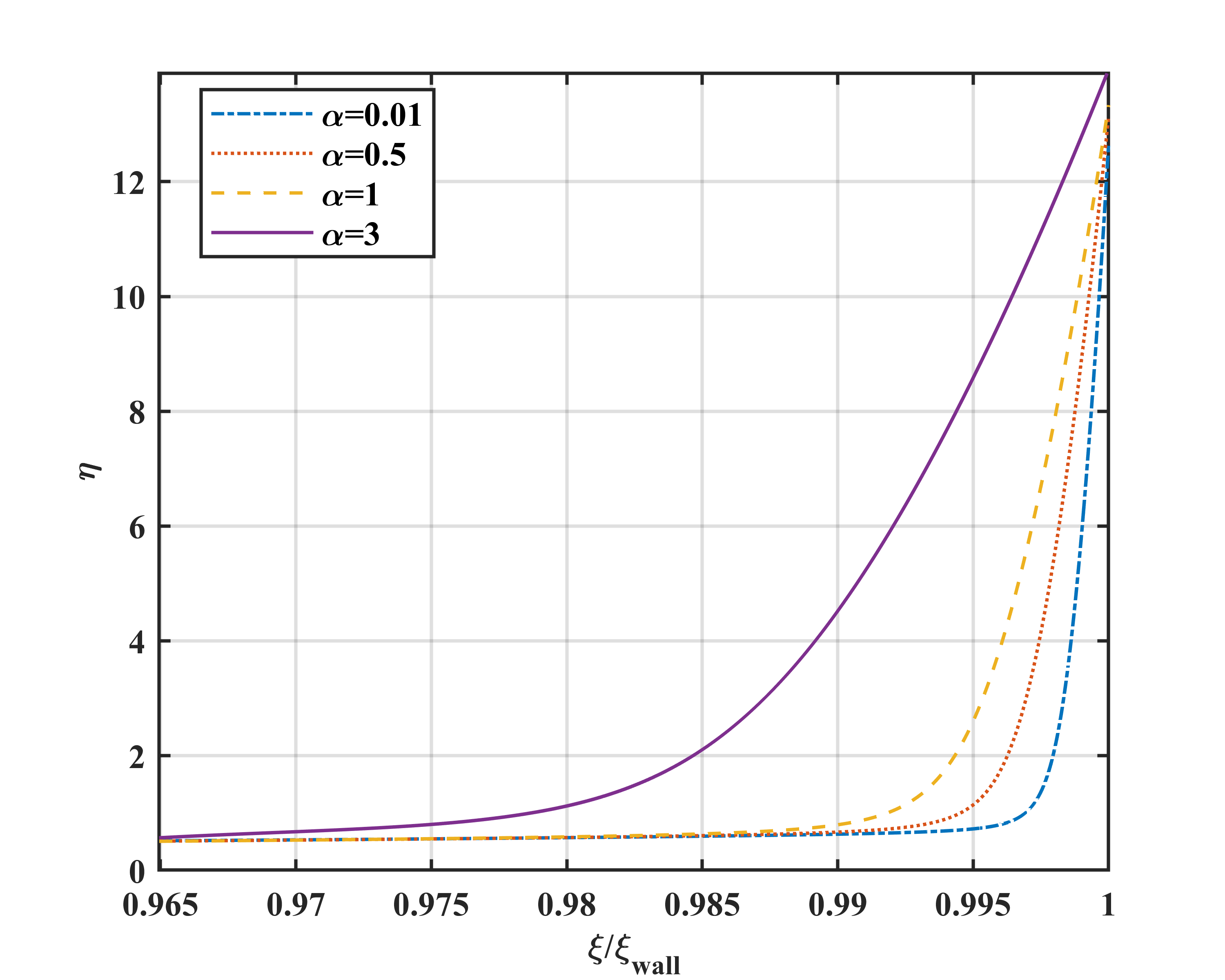}
    \caption{Electric potential variation along normalized $z$ axis for different $\alpha$ for $\theta=90$}
    \label{fig:potential_2} 
\end{figure}

\begin{figure}
    \centering
    \includegraphics[width = 8.5 cm]{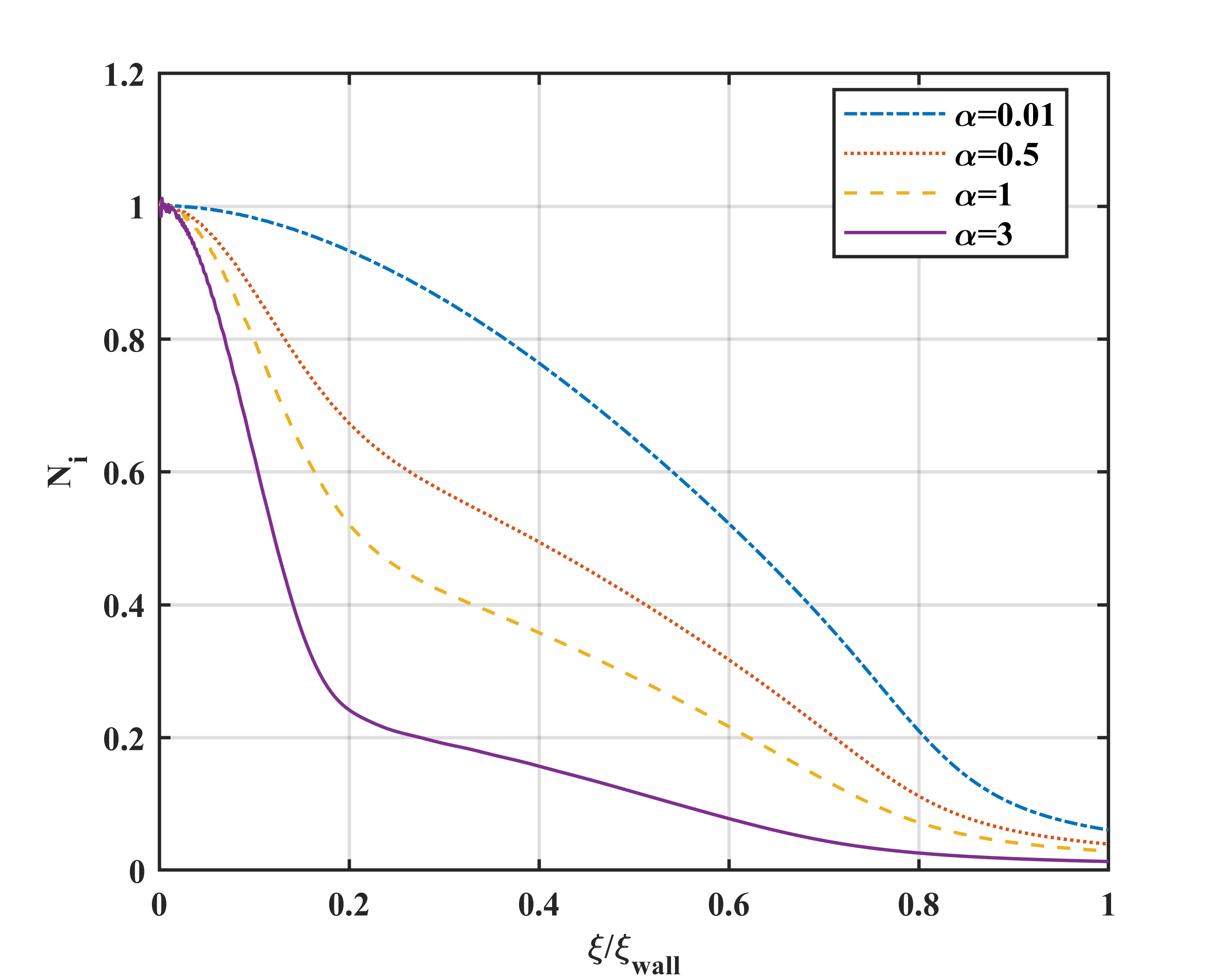}
    \caption{Normalized positive ion density variation along normalized $z$ axis.}
    \label{fig:ion density} 
\end{figure}

\begin{figure}
    \centering
    \includegraphics[width=8.5 cm]{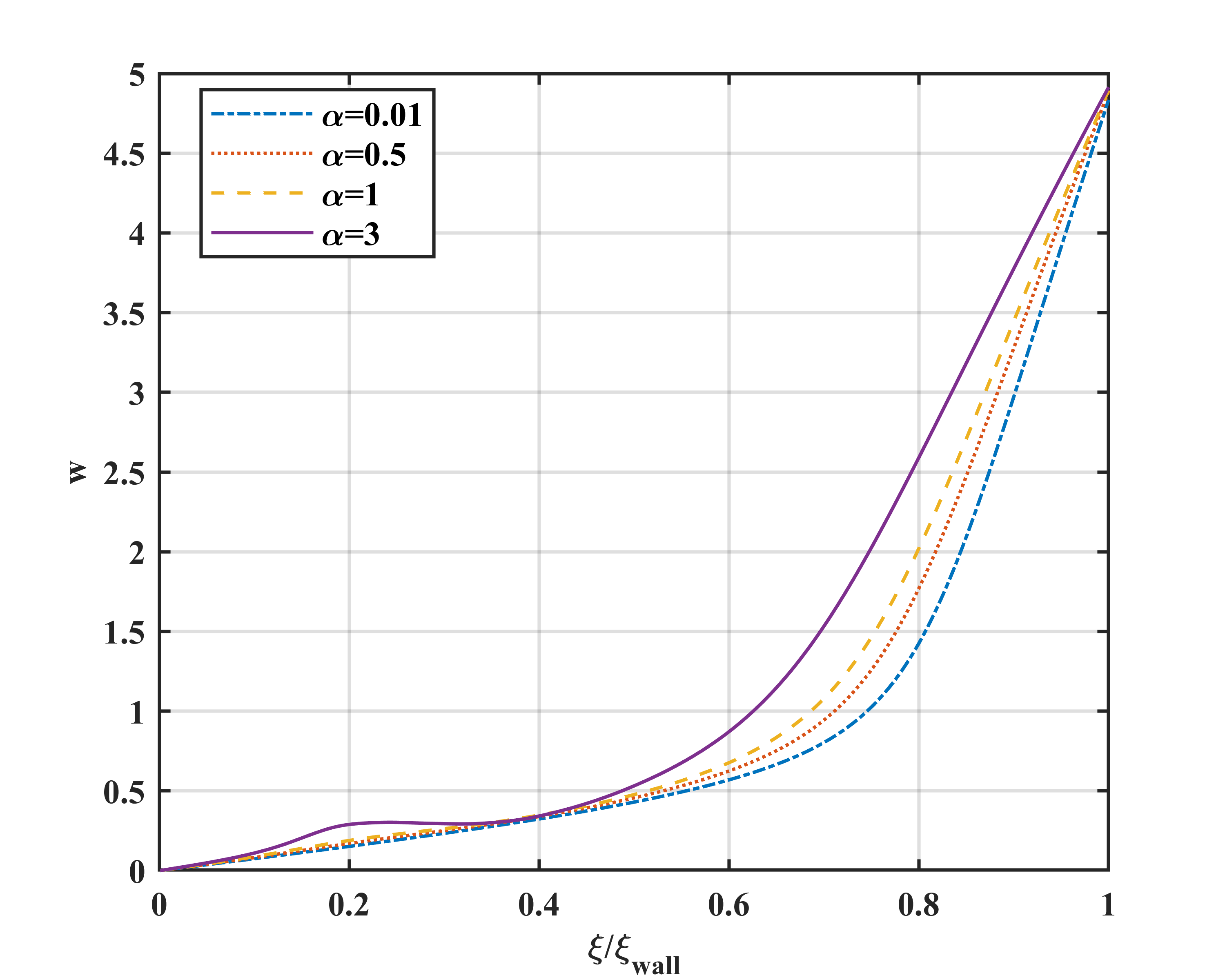}
    \caption{Normalized positive ion velocity along the $z$ direction.}
    \label{fig:velocity_w} 
\end{figure}

\begin{figure}
    \centering
    \includegraphics[width=8.5 cm]{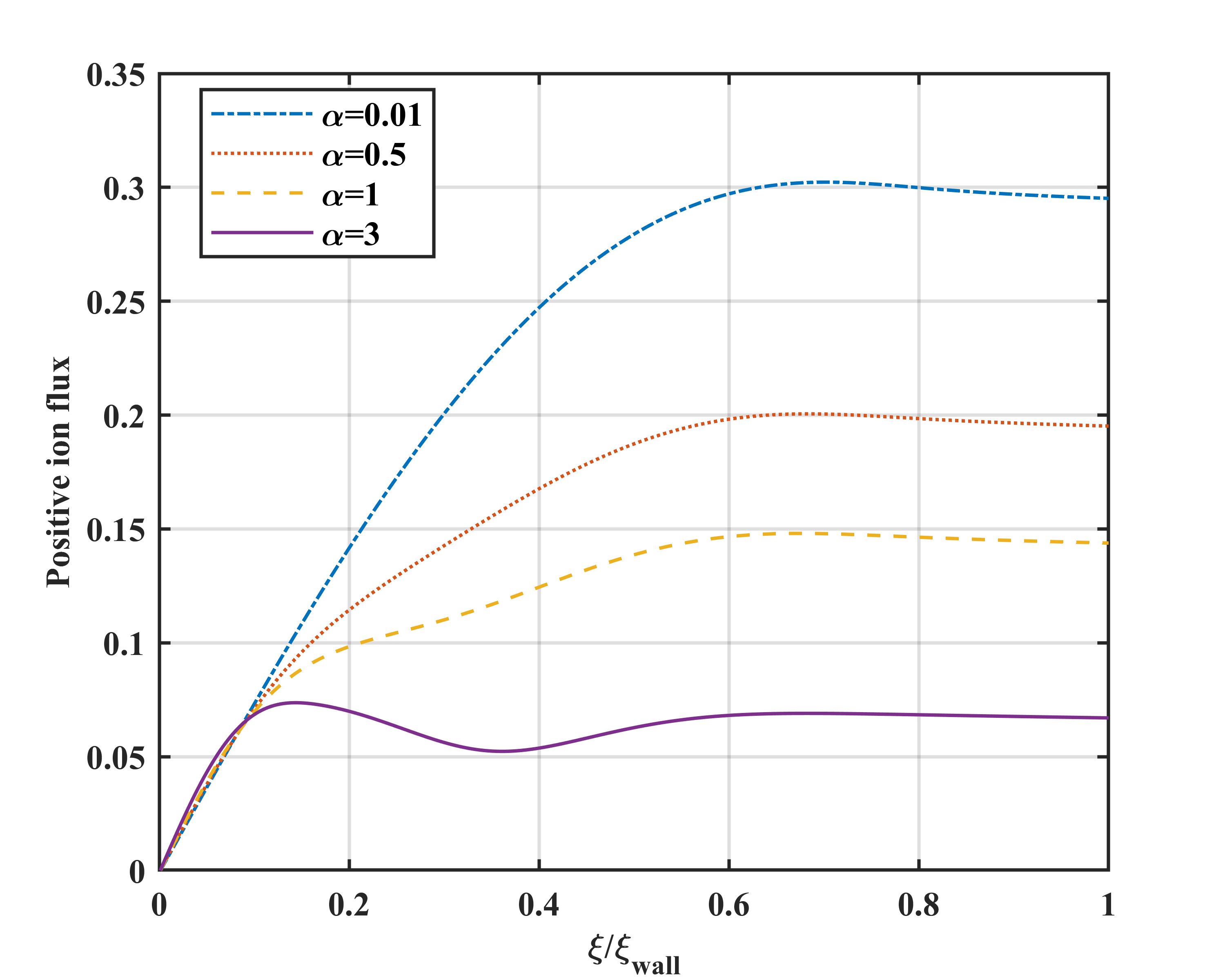}
    \caption{Normalized positive ion flux variation along normalized $z$ axis.}
    \label{fig:ion flux} 
\end{figure}

\subsection{Effect of electronegativity on sheath structure}
By varying the electronegativity, the densities of the species are studied, as shown in Fig.\ref{fig:density_1} and the obtained regimes can be broadly classified into three categories as follows
\subsubsection{Weakly Electronegative}
 For $\alpha$ = $0.01$, in this case, the density profile resembles that of an electron-ion plasma. The negative ion density is negligible in this case. 
For $\alpha$ = $0.5$, there is a nominal negative ion density, but the more substantial portion of the density is that of electrons and that of positive ions. The particle density in the sheath, in this case, is more than that of $\alpha$ = $0.01$ indicating in a thicker sheath than for  $\alpha$ = $0.01$ .
Hence, for $\alpha$ < $1$, the plasma is said to be electron-rich plasma.
 
\subsubsection{Moderately Electronegative}
For  $\alpha$ = $1$, the electrons and negative ions are seen to have equal densities. The negative ion density increases with  an increase in electronegativity. The particle density also increases which in turn further increases the sheath width.

\subsubsection{Highly Electronegative}
 In this regime, $\alpha$ = $3$, a rise in the negative ion density is observed. The major contributor to the negative species, in this case, are the negative ions. Electron density decreases significantly, but the particle density increases manifold in the sheath, thereby increasing the thickness of the sheath. 
 Therefore, for  $\alpha$ > $1$, the plasma is said to be negative ion rich plasma.\\
 
Fig.\ref{fig:space charge} represents the normalized space charge density by varying the electronegativity.The space charge profile which becomes useful in understanding the sheath thickness, it is seen that for a lower value of electronegativity, the space charge is comparatively higher than that for the higher values of $\alpha$. It signifies that a thin sheath is obtained for a lower electronegativity, which gradually increases with an increase in the electronegativity. For a lower value of $\alpha$, the wall becomes more negative, as the electrons reach the wall faster. It results in greater charge separation and consequently higher electric field, which in turn decreases the sheath width. As the electronegativity increases, this charge separation decreases, and the sheath becomes thicker. Particularly for $\alpha$ = $3$, the sheath width is maximum. With the wall being already negative due to the mobile electrons, the positive ions are attracted towards the wall. The velocity of the positive ions increases towards the wall. In the presence of the negative ions, the electron distribution widens, and the fall becomes slower as compared to the other electronegative cases. It increases the particle density in the sheath, and hence widens the sheath.\\

In Fig.\ref{fig:space_charge_2}, the sheath thickness is seen to have dependence on the angle of inclination. For a collisionless case, in magnetized scenario the sheath thickness is observed to be wider for low angle rather than for a higher value. Such an observation was earlier observed by Moulick et al. \cite{moulick2019}. However, in the present case it is observed that, the sheath is widely expanded as compared to the electrostatic case. This is little unlike the previous observation and may be induced by the presence of the negative ions. \\

The space charge profile is shown in Fig.\ref{fig:space_charge_90} for electrostatic case by varying electronegativity. It is seen that a thin sheath is formed for lower electronegativity, which increases with increase in $\alpha$. For electrostatic case , without collisions, the electric field force is the only possibility which affects the sheath formation. Since with the increase in $\alpha$, the electron population decreases, therefore, the probability of the electrons reaching the wall also decreases. It results in a less negative wall, consequently decreasing the electric field strength as reflected in the potential profile in Fig.\ref{fig:potential_2}. Thus, the widening of the space charge region attributes to the decrease in the field strength.  \\

From Fig.\ref{fig:density_1}, it is seen that the negative ion density has a sharp fall inside the sheath in comparison to the electrons. It is due to the low temperature of the negative ions. The positive ions tend to follow either the electrons or the negative ions as seen in Fig.\ref{fig:ion density}. For  $\alpha$ = $0.01$, the positive ions follow the electrons.As the electronegativity increases the positive ions initially follow the negative ions, and when the negative ions fall, they tend to follow the electrons. The negative ions affect the dynamics of the positive ions.\\ 

The normalized velocity along the $z$ direction shows an irregularity for $\alpha$ = $3$ as shown in Fig.\ref{fig:velocity_w}. In Fig.\ref{fig:ion flux},the positive ion flux also shows a deformation in the same region.The positive ion density dips at the same region as seen in Fig.\ref{fig:ion density} and its velocity along the $z$-direction increases.
Thus, the positive ion flux increases in this region. The velocity of the ions increases towards the wall which is reflected in the ion flux profile, and it becomes constant inside the sheath.\\
\begin{figure}
    \centering
    \includegraphics[width = 8.5 cm]{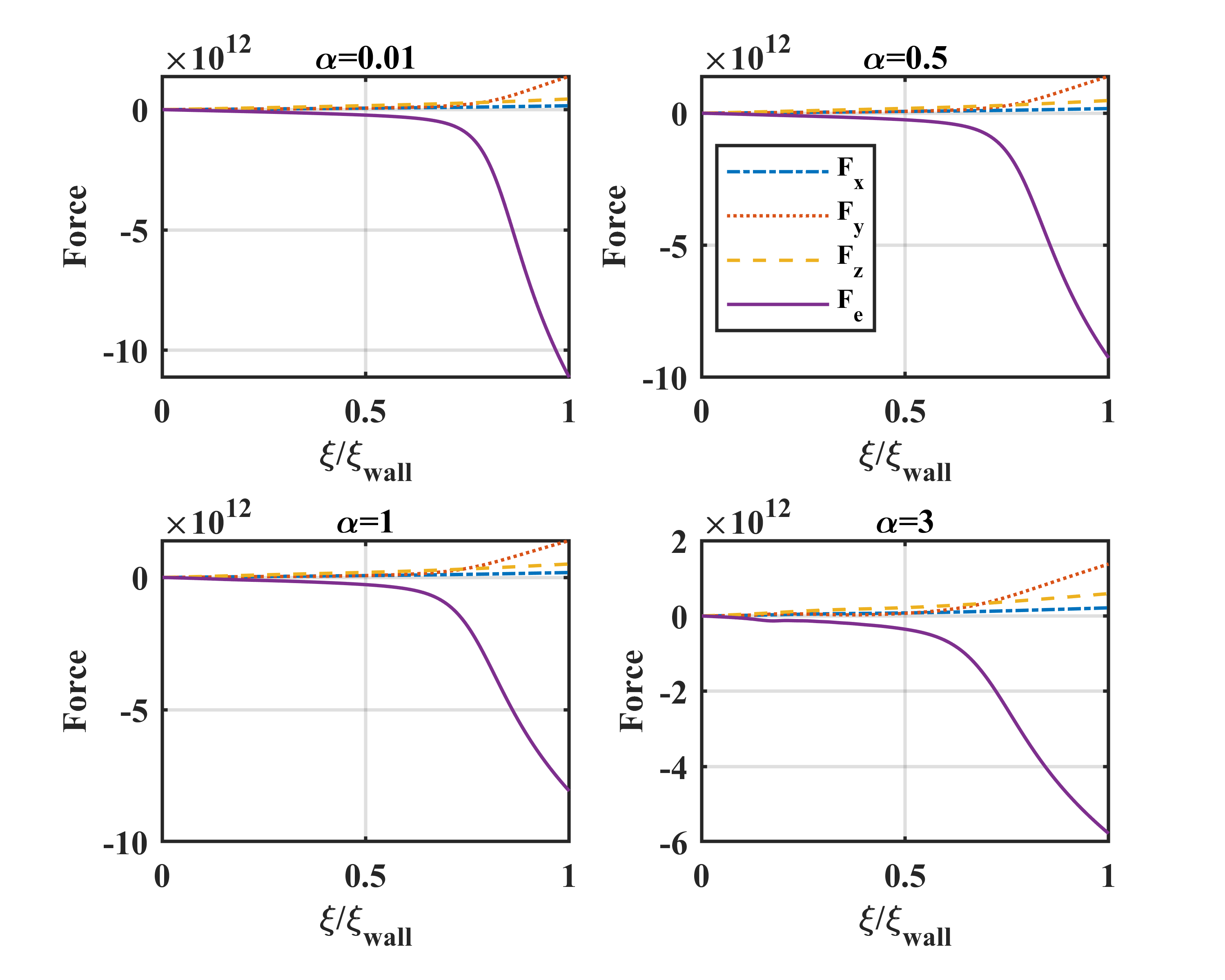}
    \caption{Normalized Lorentz force variation along normalized $z$ axis.}
    \label{fig:force} 
\end{figure}
\begin{figure}
    \centering
    \includegraphics[width = 8.5 cm]{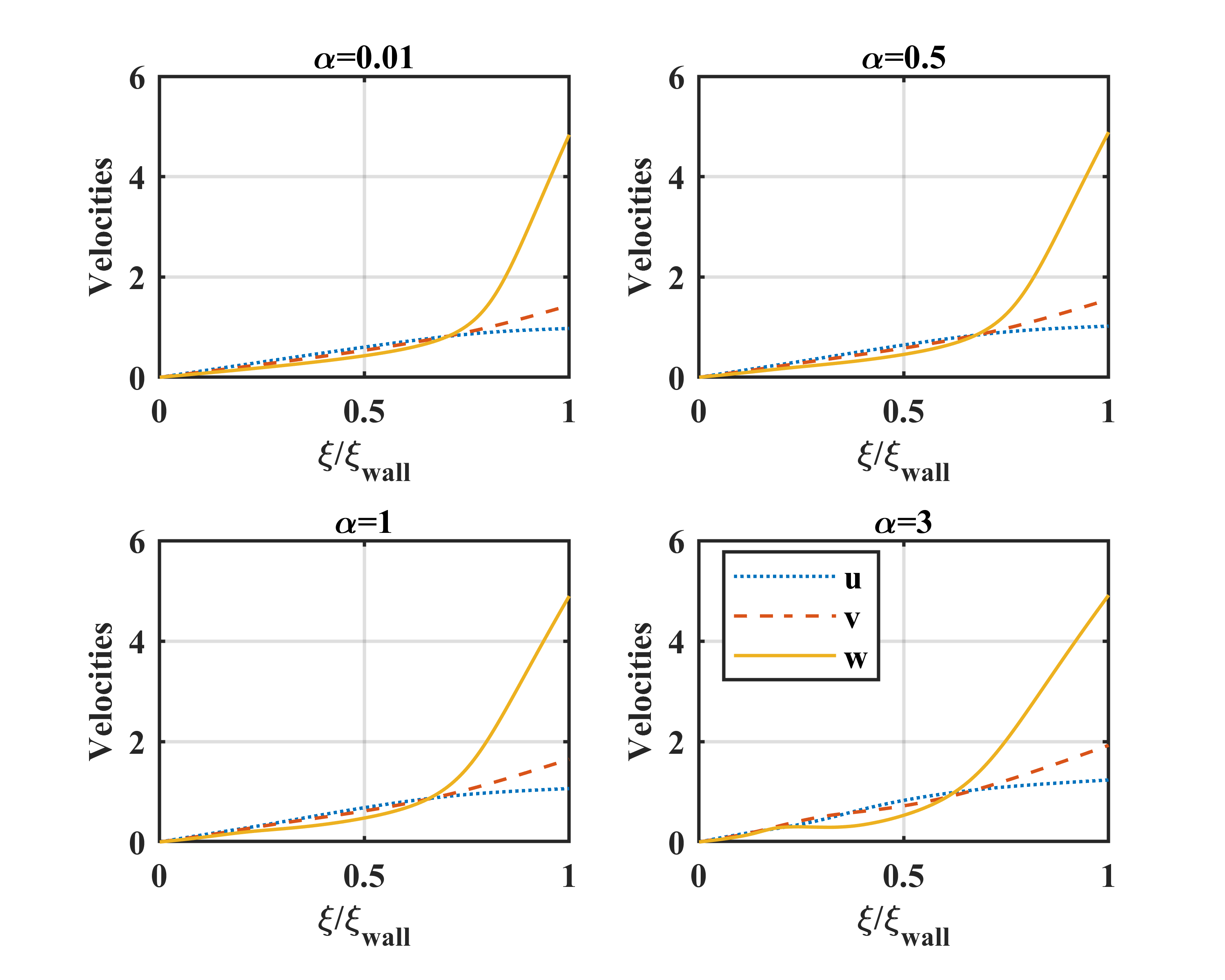}
    \caption{Normalized positive ion velocity variation along normalized $z$ axis.}
    \label{fig:velocity} 
\end{figure}
\begin{figure}
    \centering
    \includegraphics[width = 8.5 cm]{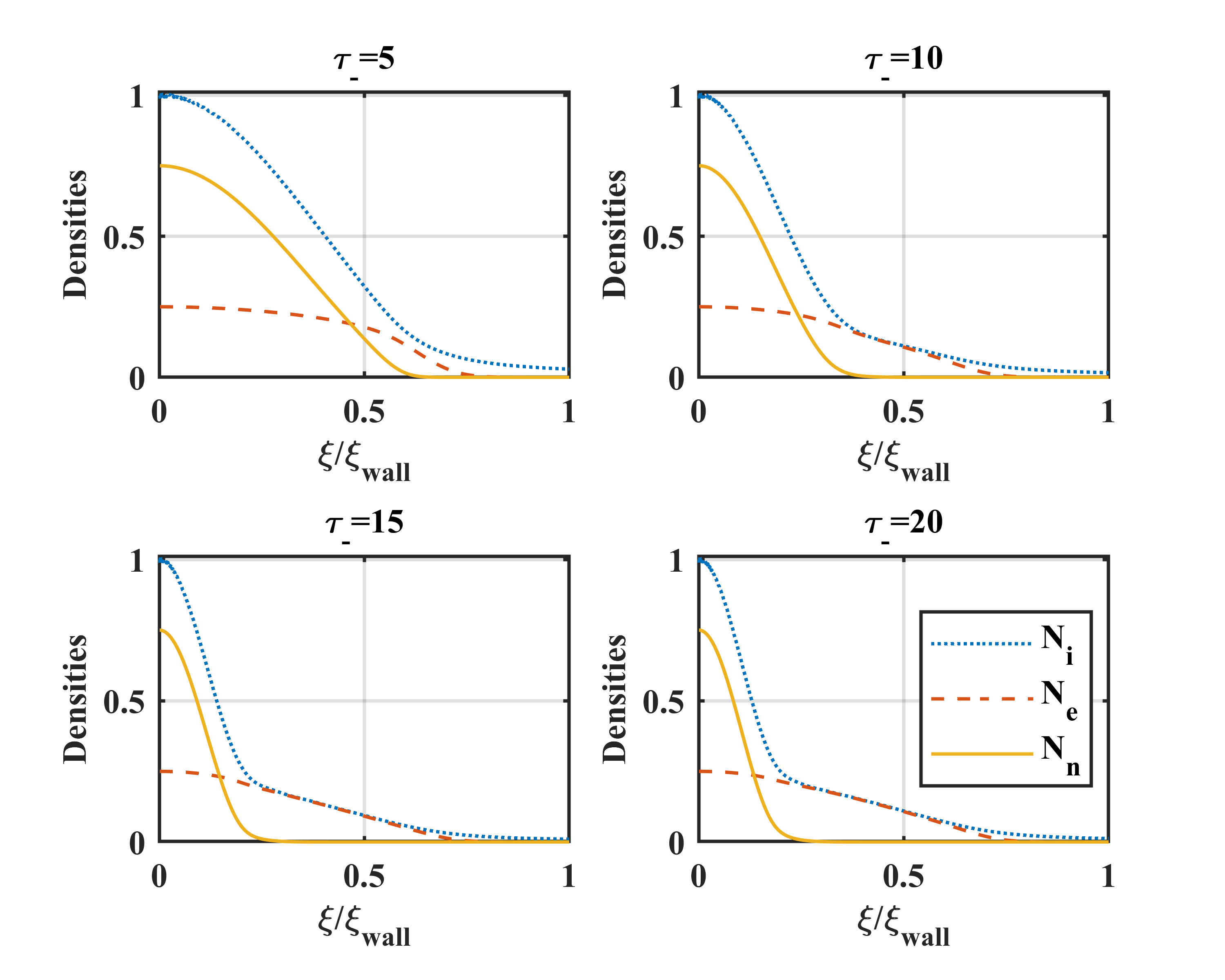}
    \caption{Normalized species density variation along normalized $z$ axis for different ${\tau_-}$ for $\alpha$ = $3$.}
    \label{fig:density_2} 
\end{figure}

The Lorentz force is shown in Fig.\ref{fig:force} for different electronegativities. It is seen that for all the cases, the magnetic force along the $y$-direction increases. $F_y$ depends on $\gamma_x$, $w$, $\gamma_z$ and $u$. The magnitude of $\gamma_x$ and $w$ are much greater than that of $\gamma_z$ and $u$. Therefore, $F_y$ increases monotonically in all cases. The magnetic force helps in guiding the positive ions to enter the sheath. On reaching the sheath edge, the magnetic force no longer plays a major role. On the contrary, the electric force now accelerates the ions towards the wall \cite{Jing}. With the increase in electronegativity, the electric force is seen to decrease. This decrease in electric force lowers the charge separation, which further increases the particle density in the sheath and therefore, increases the sheath width. For $\alpha$ = $3$, the electric force is minimum as the charge separation is the least; hence, the sheath width is maximum in this case. \\ 

\begin{figure}
    \centering
    \includegraphics[width = 8.5 cm]{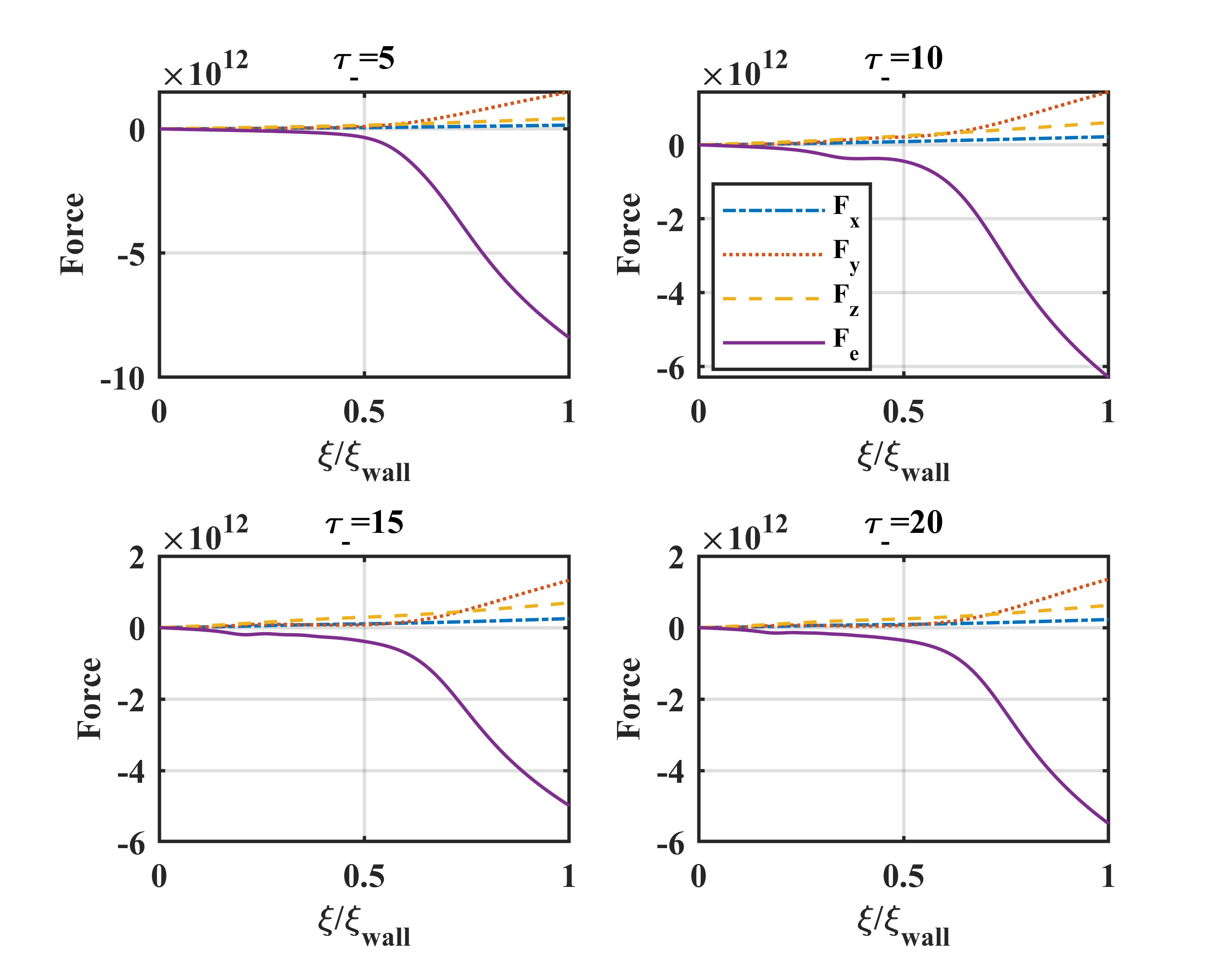}
    \caption{Normalized Lorentz force variation along normalized $z$ axis for different ${\tau_-}$ for $\alpha$ = $3$.}
    \label{fig:force_2} 
\end{figure}

With the increase in the electronegativity, it is observed that the $y$ component of velocity increases in all the cases. From equation (18) , it has been derived that the sheath formation criterion depends on the $y$ component of velocity. Fig.\ref{fig:velocity} shows that the $y$ component of velocity increases with $\alpha$ . The increase in $v_y$ is due to the $\textbf{E} \times \textbf{B}$ drift. It acts perpendicular to the $x$-$z$ plane, thereby increasing $v_y$ monotonically. \\

\subsection{Effect of temperature ratio on sheath structure}
For different values of ${\tau_-}$ , the species density is studied as shown in Fig.\ref{fig:density_2} for $\alpha$ = $3$.  As the negative ions become colder,for increasing value of ${\tau_-}$, they tend to fall faster near the source, which facilitates the electrons towards the sheath. This increases the particle density inside the sheath and thus widens the sheath.\\

The effect of ${\tau_-}$ is visible on the Lorentz force, as shown in Fig.\ref{fig:force_2}. On increasing  ${\tau_-}$, the charge separation is seen to decrease, wherein the wall becomes less negative, thereby reducing the electric force. This enhances the particle density inside the sheath, therefore increasing the sheath width.\\

\section{\label{CONCLUSIONS}CONCLUSIONS}
A numerical study on sheath using a fluid approach has been carried out for a collisionless magnetized electronegative plasma. The effect of the electronegativity and negative ion temperature on the sheath has been explored. The following conclusions can be drawn from the study \\

a) For a particular negative ion temperature, different regimes of electronegativity have been examined. It is found that for a highly electronegative plasma, the sheath width is maximum. \\

b) On increasing the electronegativity, the charge separation has been observed following an increase in the particle density inside the sheath and thereby increasing the thickness.\\

c) The dynamics of the positive ion has been affected by the presence of the negative ions. \\

d) The magnetic force along the y-direction was seen to increase with the increase in the electronegativity, and on the contrary, the electric force was lowered. With the $\textbf{E} \times \textbf{B}$ drift, acting perpendicular to the $x$-$z$ plane, the y-component of velocity has also been increasing monotonically.\\

e) For the colder negative ions, it is noted that the sheath widens. Also, they have seen to impact the electric force. The electric force decreases with the decrease in the negative ion temperature.\\

f) The present work may have useful applications in processing plasmas as well as in fusion related studies. Additionally, the role of the thermal ions can be incorporated for further study.\\

\section*{ACKNOWLEDGEMENTS}
The authors would like to acknowledge the Department of Atomic Energy, Government of India for the financial support for the present work.\\

\section*{REFERENCES}
\nocite{*}
\bibliography{aipsamp}

\end{document}